\documentclass[apj]{emulateapj}

\shorttitle{Hot gas halos around disk galaxies}

\shortauthors{Rasmussen et al.}

\begin{document}

\title{Hot gas halos around disk galaxies: \\ Confronting cosmological
  simulations with observations}


\author{Jesper Rasmussen,\altaffilmark{1,2} Jesper
  Sommer-Larsen,\altaffilmark{3,4} Kristian Pedersen,\altaffilmark{4}
  Sune Toft,\altaffilmark{5} Andrew Benson,\altaffilmark{6} Richard
  G.~Bower,\altaffilmark{7} and Lisbeth F.~Grove\altaffilmark{4}}

\altaffiltext{1}{Carnegie Observatories, 813 Santa Barbara Street,
  Pasadena, CA 91101, USA; jr@ociw.edu}

\altaffiltext{2}{Chandra Fellow}

\altaffiltext{3}{Excellence Cluster Universe, Technische
  Universit\"{a}t M\"{u}nchen, Boltzmannstr.\ 2, D-85748 Garching bei
  M\"{u}nchen, Germany}

\altaffiltext{4}{Dark Cosmology Centre, Niels Bohr Institute,
  University of Copenhagen, Juliane Maries Vej 30, DK-2100 Copenhagen,
  Denmark}

\altaffiltext{5}{European Southern Observatory, Karl-Schwarzschild-Str.\ 2, 
  D-85748 Garching bei M\"{u}nchen, Germany}

\altaffiltext{6}{Division of Physics, Mathematics, \& Astronomy,
  California Institute of Technology, Mail Code 130-33, Pasadena, CA
  91125, USA}

\altaffiltext{7}{Institute for Computational Cosmology, University of
  Durham, South Road, Durham DH1 3LE, UK}

\begin{abstract}
  Models of disk galaxy formation commonly predict the existence of an
  extended reservoir of accreted hot gas surrounding massive spirals
  at low redshift.  As a test of these models, we use X-ray and
  H$\alpha$ data of the two massive, quiescent edge-on spirals
  NGC\,5746 and NGC\,5170 to investigate the amount and origin of any
  hot gas in their halos. Contrary to our earlier claim, the {\em
  Chandra} analysis of NGC\,5746, employing more recent calibration
  data, does not reveal any significant evidence for diffuse X-ray
  emission outside the optical disk, with a $3\sigma$ upper limit to
  the halo X-ray luminosity of $4\times 10^{39}$~erg~s$^{-1}$.  An
  identical study of the less massive NGC\,5170 also fails to detect
  any extraplanar X-ray emission. By extracting hot halo properties of
  disk galaxies formed in cosmological hydrodynamical simulations, we
  compare these results to expectations for cosmological accretion of
  hot gas by spirals. For Milky Way--sized galaxies, these
  high-resolution simulations predict hot halo X-ray luminosities
  which are lower by a factor of $\sim 2$ compared to our earlier
  results reported by Toft et~al.\ (2002).  We find the new simulation
  predictions to be consistent with our observational constraints for
  both NGC\,5746 and NGC\,5170, while also confirming that the hot gas
  detected so far around more actively star-forming spirals is in
  general probably associated with stellar activity in the disk.
  Observational results on quiescent disk galaxies at the high-mass
  end are nevertheless providing powerful constraints on theoretical
  predictions, and hence on the assumed input physics in numerical
  studies of disk galaxy formation and evolution.

\end{abstract}


\section{Introduction}\label{sec,intro}

Estimates of the cosmic baryon fraction, defined as the ratio of
baryonic to total mass in the Universe, can be combined with
constraints on the integrated mass function of galaxies to infer that
most baryons in the Universe are in a hot, diffuse form at the present
epoch \citep{balo2001}.  Cosmological simulations suggest that
30--40~per~cent of all baryons reside in intergalactic filaments of
shock-heated gas with temperatures $10^5\lesssim T \lesssim 10^7$~K,
the so-called warm/hot intergalactic medium, WHIM (e.g.\
\citealt{cen1999}; \citealt{dave2001}). X-ray absorption studies have
provided observational evidence for such a component along the line of
sight towards a number of quasars (e.g.\
\citealt{trip2000,nica2005,sava2005}).

While most of the WHIM baryons are predicted to reside in structures
of low overdensity, outside the dark matter halos of individual
galaxies and groups of galaxies \citep{dave2001}, a potential
repository for some of the ``hidden'' baryons in the Universe could be
extended halos of hot ($\sim 10^6 - 10^7$ K) gas around individual
galaxies, including spirals (e.g.\ \citealt{fuku2006}). For massive
galaxies, cosmological simulations suggest that the total mass of
these ``external'' galactic baryons is comparable to that of stars and
cold gas in the galaxies themselves \citep{somm2006}. In a
cosmological context, observational support for such a scenario comes
from the angular correlations between the galaxy distribution and the
soft X-ray background, which indicate the presence of soft X-ray
emission from WHIM surrounding individual galaxies \citep{solt2006}.

The presence of extended hot gaseous halos around optically bright
elliptical galaxies is well established from X-ray observations (e.g.\
\citealt{osul2001}). The idea that disk galaxies could also be
embedded in such halos is integral to many semi-analytical models of
disk galaxy formation (e.g.,
\citealt{whit1978,whit1991,cole2000,hatt2003,bowe2006,crot2006}). In
these models, galaxy dark matter halos are assumed to grow as
predicted by spherical infall models, with gas accreting continuously
along with the dark matter.  During infall into the dark matter
potential, the gas is heated, potentially to the halo virial
temperature, subsequently cooling radiatively. If cooling is rapid, as
is the case for characteristic gas temperatures $T\la 10^6$~K and
hence relatively shallow potential wells, no accretion shock develops
outside the evolving galactic disk. In these models, present-day
spirals of total mass $M\la 10^{11}$~M$_{\sun}$ accrete gas which is
predominantly in a cold, non--X-ray emitting phase at temperatures
much lower than the virial temperature of their halo
\citep{binn1977,birn2003,binn2004,kere2005,deke2006}. This 'cold
accretion' mode would be particularly pronounced at high redshift, and
would imply that most of the halo radiation is emitted as Ly$\alpha$
emission close to the disk. Indeed, observational support for this
scenario is accumulating at both low and high redshift (e.g.,
\citealt{vand2004,frat2006,nils2006,sanc2008,smit2008}).

At higher galaxy masses and gas temperatures, the gas would emit at
X-ray wavelengths but would cool less efficiently, and is then assumed
in the models to flow in more slowly than in the fast-cooling regime.
A hot X-ray emitting halo can develop, in which the infalling gas is
eventually deposited where dictated by its angular momentum.  As in
the 'cold accretion' mode, the gas may ultimately condense to form
stars in a rotationally supported disk. In practice, thermal
instabilities could act to locally reduce the cooling time of hot halo
gas, so in addition to a classical central cooling flow, cooling may
also proceed via fragmentation of halo gas into smaller clouds
throughout the halo. The net result may be the formation of a
multi-phase halo containing warm ($T\sim 10^4$~K), pressure-supported
clouds embedded in a large-scale hot corona.  These clouds may then
rain down onto the disk, possibly akin to some of the high-velocity
H{\sc i} clouds seen around the Galaxy (e.g., \citealt{mall2004,
kauf2006,somm2006,peek2008}; but see also \citealt{frat2008}).  In the
scenario proposed by \citet{mall2004}, condensation and infall of such
clouds would be very important for fueling MW-sized galaxy disks,
although at least some numerical work suggests that this mode of
accretion is subdominant at $z=0$ \citep{somm2006,kauf2006,peek2008}.

The general picture of massive disk galaxies embedded in hot gaseous
halos has been backed by cosmological hydrodynamic simulations of
galaxy formation (e.g., \citealt{toft2002,kere2005}), thus lending
support to the general validity of the simplifying assumptions
underlying semi-analytical models. For example, the simulations
presented by \citet{toft2002} showed the X-ray luminosity $L_{\rm X}$
of hot halo gas to scale strongly with disk circular velocity $v_c$
(defined here as the rotation velocity at 2.2~times the disk
scale-length; see \citealt{somm2001} for details), with $L_{\rm X}
\propto v_c^{5-7}$, as also expected from semi-analytical models of
galaxy formation.  It is relevant to note, however, that all the
massive galaxies in the \citet{toft2002} study were formed in the old
``standard'' ($\Omega_m=1$, $\Omega_{\Lambda}=0$) cold dark matter
cosmology, with a fixed primordial metal abundance and an assumed
cosmic baryon fraction well below the currently accepted value.  Thus,
it seems timely to explore the hot halo predictions of more
sophisticated higher-resolution simulations assuming the currently
prevailing $\Lambda$CDM cosmology. This is one of the goals of this
paper.

The existing theoretical results clearly suggest that X-ray
observations could provide a useful means of mapping any material
accreting onto massive spirals at the present epoch. Direct
observational verification of the above scenario has remained elusive,
however. In a first study aimed at detecting cosmologically accreted
gas halos around spirals, \citet{bens2000} compared {\em ROSAT} X-ray
observations of three massive, highly inclined spirals to predictions
of simple cooling flow models.  The failure of these authors to detect
any hot gas around the galaxies translated into upper limits on the
halo X-ray luminosity which, for their best-constrained case of
NGC\,2841, was found to be a factor of 30 below their model
prediction. While their models were deliberately simplified versions
of general semi-analytical ones, the result persists that some
hydrodynamical simulations and simple analytical models overpredict
actual halo luminosities (\citealt{gove2004}; \citealt{bens2000} and
references therein), though the exact magnitude of this discrepancy
has yet to be established.

Extended X-ray halos {\em have} been detected around numerous
late-type star-forming spirals (see \citealt{stri2004a};
\citealt{tull2006a} and references therein). However, these galaxies
typically have at least moderately high star formation rates ($\ga
1$~M$_\odot$~yr$^{-1}$), with the detected halo X-ray emission
coinciding with extraplanar H$\alpha$ and radio continuum
emission. The standard interpretation is that these multi-phase halos
have been generated by star formation activity in the disk, with
supernovae expelling hot, X-ray emitting gas, warm and/or photoionized
H$\alpha$-emitting material, and radio synchrotron emitting cosmic ray
electrons from the disk
\citep{wang2001,wang2003,ehle2004,stri2004a,stri2004b,tull2006a}.
Unlike the expectation for gravitationally heated gas, the X-ray
luminosity of the gas in the halos of these galaxies is found to
correlate with various measures of the disk star formation rate and
with the energy input rate by supernovae (SN), but --- even when
including the diffuse X-ray luminosity of the disk --- not with
baryonic or H{\sc i} mass in the disk (\citealt{tull2006b}; see also
\citealt{sun2007}).

Diffuse X-ray emission has also been detected around massive
early-type (Sa--Sb) spirals such as NGC\,4594 (the ``Sombrero''
galaxy) \citep{li2006,li2007}. In these cases, the emission is not
obviously associated with star formation in the disk, but the total
diffuse X-ray luminosity from the disk and halo still falls well below
expectations from cosmological accretion on the basis of the
\citet{toft2002} results. It has been speculated that much of the
detected extraplanar emission in these galaxies is heated by SN~Ia
activity in the disk rather than by gravity (see, e.g.,
\citealt{wang2007}).

Summarizing, the hot halos observed so far around spirals seem to
result mainly as a consequence of stellar feedback in the disk rather
than infalling intergalactic material. A possible exception is the
Milky Way itself, a moderately star-forming galaxy for which indirect
arguments can be used to ascertain the presence of an extended
reservoir of hot gas possibly related to a gaseous halo (e.g.\
\citealt{moor1994}; \citealt{quil2001}; \citealt{semb2003};
\citealt{breg2007}; \citealt{wang2007}). Despite this, it remains
unclear whether any of this putative halo gas is actually cooling out
to accrete onto the Milky Way disk, and if so, to what extent it can
replenish the gas currently being consumed in the disk by star
formation.

Sensitive X-ray observations of carefully selected systems are
required to settle the question of whether substantial hot gas halos
associated with infalling material exist around quiescent spirals at
all. This would provide an important consistency check on the
assumptions underlying existing disk galaxy formation models.
Motivated by this, we initiated a {\em Chandra} search for hot X-ray
gas surrounding massive, quiescent disk galaxies. As part of this
program, the discovery of hot halo gas around NGC\,5746, with a total
X-ray luminosity of $\sim 4\times 10^{39}$~erg~s$^{-1}$, was reported
by \citet{pede2006}.  In the present paper, we re-visit this issue in
more detail using updated {\em Chandra} calibration data, describe our
corresponding {\em Chandra} results for the less massive galaxy
NGC\,5170, and present new results from cosmological simulations of
galaxy formation and evolution, for direct comparison to the
observations.  Contrary to the claim in \citet{pede2006}, significant
extraplanar emission surrounding NGC\,5746 is not detected in our
updated analysis.

In \S~\ref{sec,sample} we outline our target selection and the
analysis of X-ray and optical data. Results for both galaxies are
presented in \S~\ref{sec,results} and compared to predictions of
cosmological simulations and to results for generally more actively
star-forming spirals in \S~\ref{sec,comparison}. Some implications of
our observational and theoretical results are discussed in
\S~\ref{sec,discuss}, and our findings and conclusions are summarized
in \S~\ref{sec,summary}. A Hubble constant of
$H_0=73$~km~s$^{-1}$~Mpc$^{-1}$ is assumed throughout. The distances
to NGC\,5746 and NGC\,5170 as listed in NED are then 26.8 and
24.8~Mpc, respectively, with 1~arcmin corresponding to $\sim 7.7$ and
$\sim 7.1$~kpc.  Unless otherwise stated, uncertainties are reported at
the 68~per~cent confidence level.

\section{Galaxy Sample and Data Analysis}\label{sec,sample}

\subsection{Target Selection}

The primary objective of this study is to map and characterize any
diffuse X-ray gas surrounding quiescent disk galaxies. Our starting
point was therefore the simulation results of \citet{toft2002}, in
particular the result that halo X-ray luminosity $L_{\rm X}$ should
scale strongly with disk circular velocity $v_c$.  Hence, our target
sample was drawn by searching the NED and HyperLeda databases for
nearby spiral galaxies with circular velocity $v_c \ge 280$ km
s$^{-1}$. According to the results of \citet{toft2002}, this should
ensure that the target galaxies are massive enough to generate and
retain a large-scale X-ray halo detectable within reasonable {\em
Chandra} exposure times. A further advantage of large $v_c$ lies in
the fact that the efficiency with which SN outflows can escape the
disk and complicate the interpretation of our results may be expected
to decrease with increasing disk mass.

In addition, we required our targets to meet the following criteria:
i) inclination $i\ge 80\degr$, to enable a clean separation between
disk and halo gas; ii) distance $D<40$~Mpc, for similar reasons, and
to obtain a sufficiently strong X-ray signal; iii) Galactic latitude
$|b|>30\degr$, to have a low column density of absorbing foreground
Galactic hydrogen, given that any halo emission is expected to be
fairly soft and thus subject to significant absorption by intervening
material (this criterion excluded a galaxy like NGC\,2613, studied by
\citealt{li2006}); iv) the galaxies should not be a member of a known
X-ray bright group or cluster, nor show any significant signs of
interaction, so as to include only undisturbed galaxies and avoid
significant contamination from hot gas residing in an ambient
intragroup/-cluster medium; v) in order to also minimize contamination
from outflows of hot gas, the galaxies should be relatively quiescent,
i.e.\ not show evidence for significant activity related to a
starburst or an active galactic nucleus (AGN). This excluded galaxies
such as NGC\,891, which displays extraplanar X-ray, H$\alpha$, and
radio emission associated with stellar feedback activity (see
\citealt{temp2005} and references therein), and the strongly
bulge-dominated Seyfert NGC\,4594 (the 'Sombrero'), which hosts a
$\sim 10^9$~M$_\odot$ low-luminosity AGN \citep{fabb1997,bend2006}.

Imposing these strict criteria, only one galaxy remained, NGC\,5746.
We therefore relaxed the requirement on $v_c$ to include a second
target, NGC\,5170, which, although still massive, could act as a
consistency check on our analysis procedure.  NGC\,5746 is an SBb
spiral with a circular velocity $v_c = 318\pm 10$~km~s$^{-1}$
(HyperLeda database), and NGC\,5170 an Sc spiral with a significantly
lower $v_c$ of $247\pm3$~km~s$^{-1}$ \citep{kreg2004}.  We note that
our selection was not restricted to particular Hubble types, but given
the $v_c\geq 280$~km~s$^{-1}$ criterion, the selection method is
likely to bias against late-type spirals. Since such galaxies are
likely to display higher star formation activity at present than
early-type ones, this criterion is in turn likely to favor relatively
quiescent spirals, as desired. Salient parameters of the observed
galaxies are listed in Table~\ref{tab,galdata}.

\begin{table}
\caption{Salient parameters of the observed disk galaxies\label{tab,galdata}}
\begin{center}
\begin{tabular}{lcc}
  \tableline \hline
  Parameter & NGC\,5746 & NGC\,5170 \\ \hline
  $D$ (Mpc)   & 26.8 & 24.8 \\
  RA  (J2000) & $14^h44^m56\fs 00$ & $13^h29^m48\fs 83$ \\
  Dec (J2000) & $+01\degr 57\arcmin17\farcs 1$ & $-17\degr 57\arcmin59\farcs 4$
\\
  Hubble type & SBb & Sc \\
  $v_c$ (km~s$^{-1}$) &  $318\pm10$ & $247\pm3$ \\
  $D_{25}$ (arcmin) &  (6.92, 1.20) & (8.32, 1.20) \\
  $M_B$ & $-21.79$ & $-21.18$ \\
  $M_K$ & $-25.26$ & $-24.35$ \\
  $L_{\rm IR}$ (L$_\sun$) & $6.2\times 10^9$  & $3.7\times 10^9$  \\
  SFR (M$_\sun$~yr$^{-1}$) & $0.9\pm 0.2$ & $0.5\pm 0.1$ \\         
  Inclination  & $84\degr$ & $90\degr$ \\
  $N_{\rm H}$ (cm$^{-2}$) & $3.3\times 10^{20}$ & $6.9 \times 10^{20}$ \\
  {\em Chandra} obs.\ ID   & 3929   & 3928 \\
  {\em Chandra} obs.\ date & 2003--04--11 & 2003--05--18\\
  {\em Chandra} exp.\ (ks) & 35.7 & 31.9 \\
  \tableline
\end{tabular}
\tablecomments{Distances, positions, and magnitudes taken from the
  NASA/IPAC Extragalactic Database (NED). Disk circular velocity
  $v_c$, $D_{25}$ (major and minor diameter of the ellipse outlining a
  $B$--band isophotal level of 25~mag~arcsec$^{-2}$), and inclination
  from the HyperLeda database.  $L_{\rm IR}$ is the total
  (8--1000~$\mu$m) infrared luminosity, based on {\em IRAS} fluxes and
  the relation of \citet{sand1996}, and SFR the estimated star
  formation rates (see text for details).  Absorbing column density
  $N_{\rm H}$ is the Galactic value from \citet{dick1990}.}
\end{center}
\end{table}

As the goal of this study is to constrain the amount of {\em
infalling} hot halo gas, the requirement of low levels of disk
activity is particularly important for ensuring minimal contamination
of the extraplanar X-ray flux from {\em outflowing} hot gas. In terms
of their star formation rate (SFR), both galaxies are indeed fairly
quiescent, as already evidenced by the fact that none of them is
included in the revised catalog of bright {\em IRAS} galaxies
\citep{sand2003}, despite their close proximity and large mass. Their
{\em IRAS} 60 and 100~$\micron$ fluxes $S_{60}$ and $S_{100}$, taken
from \citet{mosh1990} for NGC\,5746 and \citet{rice1988} for
NGC\,5170, translate into global SFRs of $0.9\pm 0.2$ (NGC\,5746) and
$0.5\pm 0.1$~M$_{\sun}$~yr$^{-1}$ (NGC\,5170), using the relations of
\citet{sand1996} and \citet{kenn1998}.  This is equivalent to specific
(area-normalized) SFRs of $\sim 4\times 10^{-4}$ (NGC\,5746) and $\sim
2 \times 10^{-4}$~M$_{\sun}$~yr$^{-1}$~kpc$^{-2}$ (NGC\,5170),
assuming a circular disk of radius equal to the semi-major axis of
$D_{25}$.  For comparison, the moderately quiescent Milky-Way disk
displays an SFR of $\sim 2$~M$_{\sun}$~yr$^{-1}$ (see e.g.\
\citealt{casu2004}), corresponding to $\sim 3\times
10^{-3}$~M$_{\sun}$~yr$^{-1}$~kpc$^{-2}$. Both galaxies furthermore
display low far-infrared 'temperatures' of $S_{60}/S_{100}\approx
0.15$ (NGC\,5746) and $0.22$ (NGC\,5170), and their far-infrared
luminosities $L_{\rm FIR}$ imply "mass-normalized" star formation
rates $L_{\rm FIR}/L_{\rm B} \approx 0.06$, further indicative of low
star formation activity (see e.g.\ \citealt{read2001} for details).

From a {\em Chandra} study of extraplanar X-ray emission around nearby
disk galaxies covering a range of morphologies and star formation
rates, \citet{stri2004b} derived a criterion for gas blow-out by
supernovae based on the observed properties of the interstellar medium
in typical spirals.  They find on theoretical grounds that in order to
have a collection of near-simultaneous SN expel gas from the disk, the
requirement $F_{{\rm SN,FIR,}D_{25}} \ga 25$~SN~Myr$^{-1}$~kpc$^{-2}$
must be satisfied.  In their notation, $F_{{\rm SN,FIR,}D_{25}}$ is
the rate ${\cal R}_{\rm SN} = 0.2L_{\rm IR}/10^{11}$~L$_{\sun}$ of
core-collapse SN per disk ``area'' $D_{25}^2$.  Indeed, for the eight
out of ten spirals in their sample with detected extraplanar X-ray
emission, they find $F_{{\rm SN,FIR,}D_{25}}\ga
40$~SN~Myr$^{-1}$~kpc$^{-2}$.  Figure~\ref{fig,strickland} compares
the core-collapse SN rates of the galaxies in the
\citet{stri2004a,stri2004b} sample to those inferred for our target
galaxies using the same methods. The additional galaxies of
\citet{tull2006a} from a comparable {\em XMM-Newton} study have also
been plotted. Both our targets clearly lie in a regime in which
ejection of detectable hot gas by SN would not be anticipated. While
these results neglect the contribution from type~Ia SNe, low specific
SN~Ia rates of $\approx 4.7$ (NGC\,5746) and
$3.3$~SN~Myr$^{-1}$~kpc$^{-2}$ (NGC\,5170) can be estimated, using the
prescription adopted by \citet{mann2005} and the $K$- and $B$-band
magnitudes listed in Table~\ref{tab,galdata}.  Adding these to the
results in Fig.~\ref{fig,strickland} clearly still places both
galaxies well below the blowout criterion derived by
\citet{stri2004b}. These results suggest that any extraplanar emission
surrounding our targets will be largely uncontaminated by emission
from supernova-driven outflows of hot gas, enabling reliable
constraints on the physical properties of any accreting material.

\begin{figure}
\begin{center}
\epsscale{1.15}
\plotone{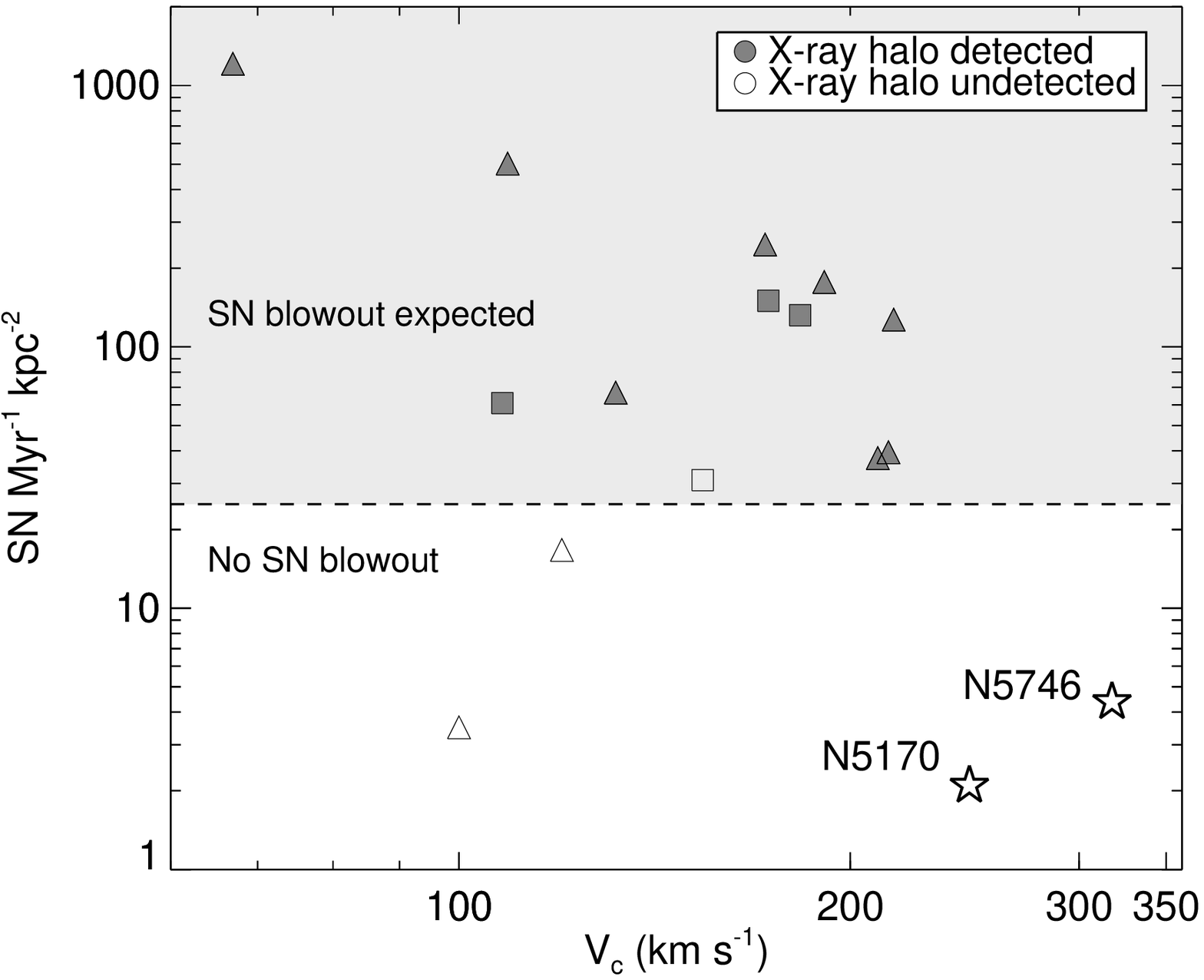}
\end{center}
\figcaption{Comparison of the specific core-collapse SN rates
    estimated for our target galaxies (stars) to those in the disk
    galaxy sample of \citet{stri2004a,stri2004b} (triangles) and
    \citet{tull2006a} (squares). Galaxies with detected extraplanar
    X-ray emission (in all cases attributed to stellar feedback) are
    represented by filled symbols. The dashed line marks the expected
    critical SN rate, $\sim 25$~SN~Myr$^{-1}$~kpc$^{-2}$, above which
    SN are expectedly capable of expelling gas from the disk.
    \label{fig,strickland}}
\end{figure}

\subsection{Observations and Analysis}\label{sec,obs}

Both disk galaxies were observed by {\it Chandra} with the ACIS-I
array as aimpoint and with the CCD's at a temperature of
$-120^\circ$~C. Data were telemetered in Very Faint mode which allows
for superior background suppression relative to standard Faint
mode. To exploit this, the data were reprocessed and background
screened using {\sc ciao} v3.4 with CALDB 3.4.1. Bad pixels were
screened out using the bad pixel map provided by the reduction
pipeline, and remaining events were grade filtered, excluding {\em
ASCA} grades 1, 5, and 7.  Periods of high background on the ACIS-I
chips were filtered using $3\sigma$ clipping of the 0.3--12~keV
lightcurves extracted in off-source regions in 259-s bins. Resulting
lightcurves showed no strong flaring periods, leaving a total of
35.7~ks (NGC\,5746) and 31.9~ks (NGC\,5170) of cleaned exposure time.

Point source searches were carried out with the {\sc ciao} task
`wavdetect' using a range of scales and detection thresholds, and
results were combined. A total of 126 (NGC\,5746) and 111 (NGC\,5170)
point sources were detected in the two observations, of which 20 and
17 sources, respectively, are located inside the $D_{25}$ ellipses.
Point source extents were quantified using the $4\sigma$ detection
ellipses from `wavdetect', and these regions were masked out in the
analysis of diffuse emission. Further details of the X-ray analysis
are described below.
 
\begin{figure*}
\begin{center}
\epsscale{1.}  
\plotone{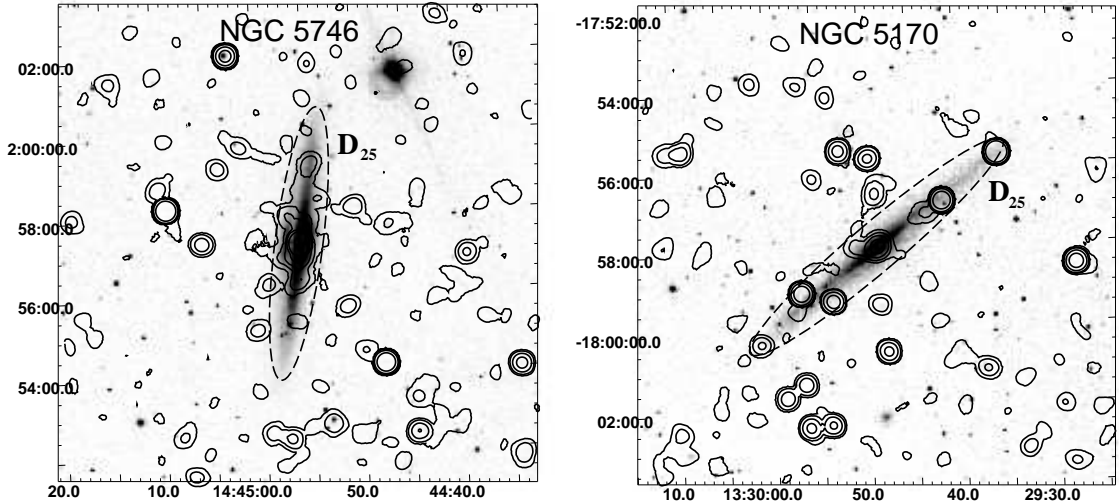}
\end{center}
\figcaption{0.3--2~keV exposure-corrected X-ray contours overlayed on
    our H$\alpha$ images of both galaxies. Contours have been smoothed
    with a Gaussian of width $\sigma=20\arcsec$, and are
    logarithmically spaced over a decade beginning at 1.5 times the
    background level. The $D_{25}$ ellipse is outlined by a dashed
    contour in each case.
    \label{fig,overlays}}
\end{figure*}

In addition to the X-ray data, we also obtained H$\alpha$ images of
both galaxies with the Danish 1.54-m telescope at La Silla, Chile, for
total exposures of 120 (NGC\,5746) and 160~min (NGC\,5170). In both
cases, 20~min $R$-band images were also taken for subsequent
continuum-subtraction.  The frames were bias-subtracted, flat-fielded,
and median-combined using standard {\sc iraf} procedures.  Using the
$R$-band image of each galaxy, we estimated and subtracted the
continuum emission contribution to the H$\alpha$ image as follows.
The two combined images were normalized to the same exposure time, and
the $R$-band image was smoothed with a Gaussian kernel to match the
broader point spread function of the H$\alpha$ image. The smoothed
$R$-band image was then scaled to match the narrower width of the
H$\alpha$ filter and subtracted from the H$\alpha$ image.  For
NGC\,5746, the background in the resulting continuum-subtracted
H$\alpha$ image had a large-scale gradient caused by stray light from
a star just outside the field of view. We modeled this using the {\sc
SExtractor} software \citep{bert1996} with options set to save a full
resolution interpolated background map. Parameters were optimized to
generate a map sufficiently fine to represent large scale variations,
and sufficiently coarse so as not to include local non-background
structures (such as an H$\alpha$ halo around the galaxy). The
resulting background map was then subtracted from the H$\alpha$ image.

\section{Results}\label{sec,results}

\subsection{X-ray and H$\alpha$ Emission}

In order to aid the search for diffuse X-ray emission around either
galaxy, smoothed, exposure-corrected 0.3--2~keV images were created
for both using two separate methods. One involved smoothing the images
with a simple Gaussian kernel, while in the other we generated
background-subtracted images following the procedure outlined in
\citet{rasm2006}. In either case, the resulting images were
exposure-corrected using similarly smoothed exposure maps. The overall
impression conveyed by these images is qualitatively similar,
suggesting very little diffuse emission beyond the optical disk of
either galaxy. Given the expected weakness of any halo signal, we here
only discuss the simple Gaussian smoothed images, so as to accommodate
any concerns that adaptive smoothing may introduce spurious features
for these low--S/N data. Fig.~\ref{fig,overlays} shows our H$\alpha$
images of the central $12\times 12$~arcmin region around each galaxy,
with the Gaussian smoothed X-ray contours overlayed, and with the
optical extent of the galaxies outlined by their respective $D_{25}$
ellipses. Clearly, both diffuse and point-like X-ray emission is
detected within the optical disks of both galaxies, but there is no
immediate indication of any diffuse X-ray or H$\alpha$ emission
extending well beyond the disks.

In order to perform a more quantitative search for diffuse X-ray
emission around the galaxies, and confirm the overall impression
conveyed by Fig.~\ref{fig,overlays}, we generated 0.3--2~keV surface
brightness profiles of the unsmoothed, exposure-corrected emission.
Point sources, chip gaps, and emission inside $D_{25}$ were masked
out.  Radial profiles extracted within $r=40$~kpc ($r\approx 5$~arcmin
for NGC\,5746) are shown in Figure~\ref{fig,surfbright}, binned into
ten equal-sized radial bins for improved S/N, and with the background
level in the data evaluated from a surrounding ($r=5.5$--7.5~arcmin)
point source--excised annulus.

These profiles support the conclusion that no significant detectable
X-ray emission surrounds either galaxy on large scales. We confirmed
that this result does not depend on the specific choice of radial
binning scheme.  While the NGC\,5170 profile shows emission outside
$D_{25}$ which is everywhere consistent with a constant level equal to
the estimated mean background level, the profile for NGC\,5746 does
show some deviations from a uniform level. Excess emission is detected
in the radial bin immediately outside the optical disk at $2.2\sigma$
significance, suggesting the presence of faint extraplanar emission at
$\sim 5$--10~kpc from the galactic center. We investigate this in more
detail in the inset in Fig.~\ref{fig,surfbright}a, which displays the
NGC\,5746 profile extracted in bins perpendicular to the disk.  This
plot confirms that the excess is not an artifact of the adopted
binning scheme, while also indicating that the excess may be largely
confined to the eastern side of the disk. However, since the signal is
only nominally significant at the $\sim 2\sigma$ level and is
comprised of only 15~net photons, we do not consider it robust
evidence of diffuse halo emission associated with NGC\,5746.  It is
also worth remarking that a comparably significant {\em deficit} of
emission is seen at larger radii ($r\approx 3.5$--4~arcmin),
suggesting that the excess may be a spurious feature introduced by
local fluctuations in the background level, perhaps caused by the
presence of point sources just below our detection limit.

\begin{figure}
\begin{center}
\epsscale{.95}
\plotone{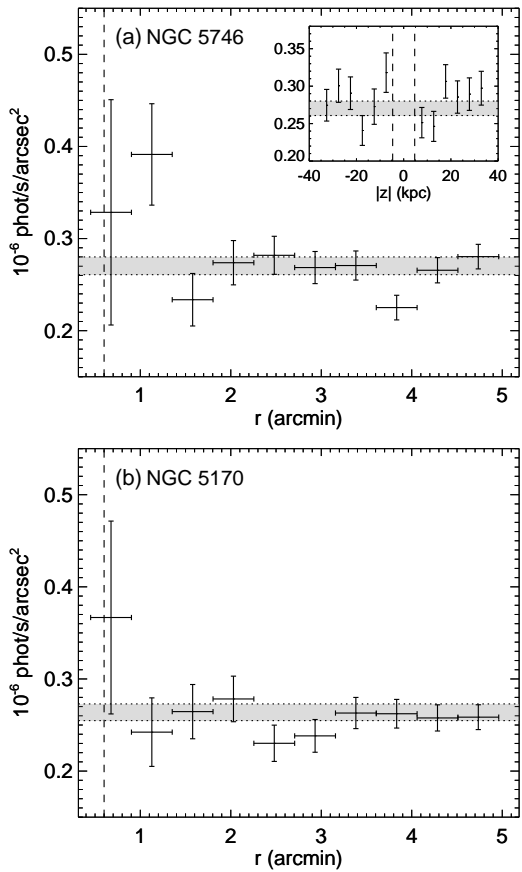}
\end{center}
\figcaption{0.3--2~keV exposure-corrected radial surface brightness
    profiles of (a) NGC\,5746 and (b) NGC\,5170. Dashed vertical lines
    outline the maximum extent of the $D_{25}$ ellipse minor axis, and
    shaded regions mark the 90\% confidence uncertainties on the local
    background level. Point sources, chip gaps, and emission within
    the $D_{25}$ ellipse itself have been masked out. Inset in (a)
    shows the NGC\,5746 profile (same units, with east to the left)
    extracted perpendicular to the disk, with $D_{25}$ again masked
    out.
    \label{fig,surfbright}}
\end{figure}

Inside the circular aperture underlying Fig.~\ref{fig,surfbright},
i.e.\ with the $D_{25}$ ellipse excluded, we detect a total of 38~net
photons in 0.3--2~keV for NGC\,5746, corresponding to emission at just
$0.8\sigma$ above the background. The corresponding result for
NGC\,5170 shows a {\em deficit} of 101~photons (significant at
$2.1\sigma$) relative to the surrounding background. Qualitatively
similar results are obtained in narrower energy bands such as
0.5--1.5~keV, in which the number of source photons is obviously lower
but the ACIS-I S/N from any absorbed $T\sim 0.2-0.3$~keV plasma should
be higher.  Hence, the immediate conclusion is that no statistically
significant diffuse X-ray emission is detected outside the optical
disk of either galaxy on these scales.

In other cases where diffuse X-ray emission has been reported around
spirals, this emission is typically accompanied by extended H$\alpha$
and radio halos, with similar overall morphologies (e.g.,
\citealt{stri2004a, tull2006a}).  Although there is no indication in
Fig.~\ref{fig,overlays} of significant extraplanar H$\alpha$ emission
around NGC\,5746, we investigate this possibility in more detail in
Fig.~\ref{fig,Hasurf}, where we show the surface brightness profile of
the H$\alpha$ emission perpendicular to the disk of this galaxy.  As
can be seen, there is no evidence for significant amounts of H$\alpha$
gas outside the optical extent of the disk. Note that the dip in the
H$\alpha$ profile 10~arcsec east of the $D_{25}$ center is due to a
dust lane also visible in broad-band optical images. For NGC\,5170, a
similar study has been performed by \citet{ross2000}, who find no
evidence for extraplanar diffuse H$\alpha$ emission surrounding this
galaxy in an H$\alpha$ exposure of similar depth to ours. There is
also no indication of significant extraplanar {\em radio} emission
around either galaxy in the public NRAO VLA Sky Survey (NVSS) 1.4~GHz
data of \citet{cond1998}. For NGC\,5746, this conclusion is further
supported by the NVSS flux density profile extracted from these data,
as shown in Fig.~\ref{fig,Hasurf}b. These results provide further
testimony to the low star formation activity of both our galaxies, and
support the assumption based on Fig.~\ref{fig,strickland} that our
constraints on X-ray emission associated with any gas accreting onto
these galaxies are not subject to serious contamination by emission
from disk outflows.

\begin{figure*}
\begin{center}
\epsscale{1.} 
\plotone{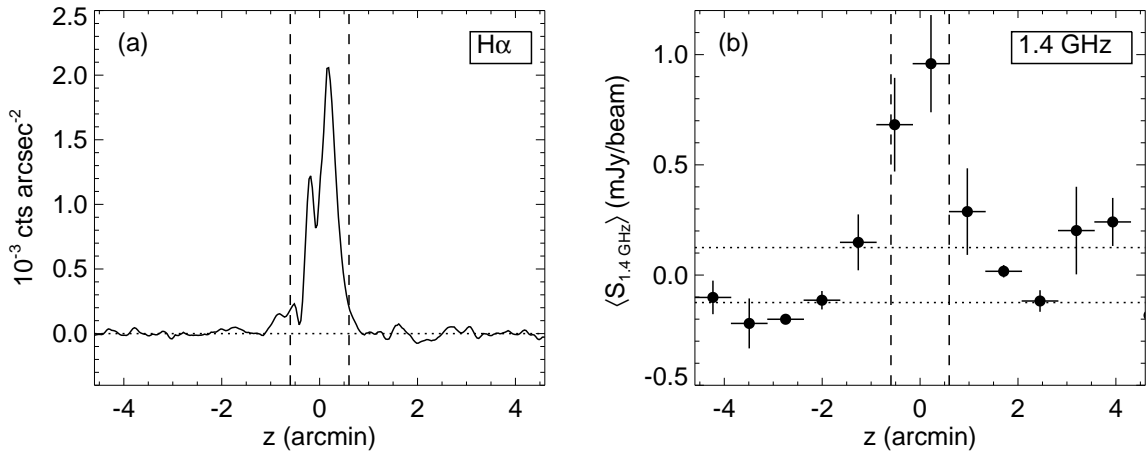}
\end{center}
\figcaption{(a) Continuum-subtracted H$\alpha$ surface brightness
  profile of NGC\,5746, extracted perpendicular to the disk within the
  same aperture as that underlying the inset in
  Fig.~\ref{fig,surfbright}a (again with east to the left).  Bright
  stars have been masked out. (b) Corresponding mean flux density of
  1.4~GHz continuum emission within the same aperture, in bins
  corresponding to the VLA beam size of 45~arcsec (FWHM). Bright point
  sources have been masked out. Uncertainties represent the standard
  deviation within each bin, and dotted lines outline the $1\sigma$
  errors on the noise level.  Dashed vertical lines are as in
  Fig.~\ref{fig,surfbright}.
\label{fig,Hasurf}}
\end{figure*}

\subsection{Differences with Respect to \cite{pede2006}}

The absence of significant detectable diffuse emission around
NGC\,5746 contrasts with the earlier result of \cite{pede2006} in
which a $4\sigma$ detection of halo emission was claimed, with a
0.3--2~keV luminosity $L_{\rm X} = 4.4^{+3.0}_{-1.5}\times
10^{39}$~erg~s$^{-1}$ and a plasma temperature of
$T=0.56^{+0.18}_{-0.20}$~keV.  Although the modest significance of the
detection renders this result tentative rather than conclusive, the
difference with respect to our present conclusion demands an attempt
at an explanation. We speculate that this discrepancy arises due to a
change in the calibration of the time-dependent degradation of the
low-energy ACIS CCD quantum efficiency, a possibility also advocated
by \citet{wang2007}.  This degradation is believed to be caused by a
build-up of molecular ice on the {\em Chandra}/ACIS optical blocking
filters or
CCD's\footnote[1]{http://cxc.harvard.edu/cal/Acis/Cal\_prods/qeDeg/},
an effect which was known but still not well understood at the time of
our original analysis in \citet{pede2006}.

The properties of the contaminant has remained the focus of continued
measurements and is considerably better understood at present. In
particular, a model of the {\em spatial} dependence of the contaminant
was introduced with the calibration data in CALDB v.\ 3.0.  However,
the analysis of \citet{pede2006} was performed using CALDB v.\ 2.26,
with which the contamination could only be appropriately corrected for
in the case of an on-axis source.  As contamination is observed to
increase towards the edges of the ACIS array, not correctly accounting
for this spatial variation will introduce a spurious signal in
exposure-corrected low-energy ACIS-I images, appearing as faint
diffuse emission whose surface brightness decreases with distance from
the CCD aimpoint. It will also lead one to underestimate the soft
X-ray background at large off-axis angles.

We conducted a few tests to explore whether this effect could account
for the halo signal detected by \citet{pede2006}. At a photon energy
$E=0.7$~keV, close to the flux peak energy of a $T\approx 0.6$~keV
thermal plasma (the halo temperature inferred by \citealt{pede2006}),
the additional optical depth towards the edges of the CCD arrays can
be 0.3--0.5 for data taken around the time of our observations (see
previous footnote), implying an apparent reduction in surface
brightness at large off-axis angles to 60--75\% of the on-axis
value. The best-fit (halo plus background) surface brightness model
reported by \citet{pede2006} in fact declines at large distances from
the NGC\,5746 disk to about 75\% of the value seen immediately outside
the optical disk (at $|z|\approx 1.5\arcmin$), consistent with the
expectation for the spatial variation of the contaminant.

As a further test of this hypothesis, we re-consider the spectral
analysis described by \citet{pede2006}, in which a double-subtraction
approach was employed for the extraction of a background
spectrum. This involves extracting a background spectrum for the same
detector region as the source region, using blank-sky background data
prepared in a similar manner to the source data. This procedure is
then repeated for a large-radius annulus, in order to obtain a
residual background spectrum that helps to account for any difference
in soft X-ray background between source- and blank-sky data. This
residual spectrum is then added to the source-region blank-sky
spectrum to produce a composite background spectrum.  We here repeated
the exact same procedure as in \citet{pede2006}, while employing the
more recent calibration files included with CALDB 3.4.1. In order to
mimic the effect of underestimating the soft X-ray background at large
radii, as implied by our hypothesis, the normalization of the soft
residual spectrum was artificially scaled down so as to achieve $\sim
200$ net counts in the source region as reported by \citet{pede2006}.
The resulting background-subtracted source spectrum was then fitted
precisely as in our original analysis, yielding a best-fit temperature
and unabsorbed 0.3--2~keV luminosity of $T=0.56^{+0.19}_{-0.15}$~keV
and $L_{\rm X}= 4.0^{+1.5}_{-1.2}\times 10^{39}$~erg~s$^{-1}$,
respectively, with $\chi^2=7.1$ for 9 degrees of freedom. These values
are strikingly similar to those obtained by \citet{pede2006}.

In summary, both the spatial and spectral properties of the diffuse
X-ray halo around NGC\,5746 reported by \citet{pede2006} are
consistent with the hypothesis that the apparent halo signal was an
instrumental artifact caused by the spatial variation in the
degradation of the low-energy ACIS quantum efficiency, an effect which
our original analysis could not take into account.

The question then arises as to why this calibration problem did not
lead to a similar spurious detection of a hot halo around NGC\,5170,
given that the two galaxies were observed with ACIS-I within one month
of each other and with comparable exposure times. The low
signal-to-noise ratio outside the disk in both data sets, including
that of the original halo ``detection'' around NGC\,5746, precludes
robust conclusions in this regard however, so we can only offer some
general considerations. We note that we find no statistically
significant differences in the background level between the two data
sets in any of a dozen different energy bands or large-radius
background regions considered, so global background differences are
unlikely to play a major role. The slightly different data preparation
criteria applied in our updated analysis could have some impact
though, changes in the calibration data aside.  For example, our
lightcurve screening employs somewhat larger time bins, and has
removed a previously undetected minor flare at the end of the
NGC\,5170 observation. Our current point source search is also
slightly more sophisticated, and has resulted in the identification of
ten additional point-like sources in the NGC\,5170 data (but none such
for NGC\,5746), of which seven are outside the disk. In combination
with the slightly shorter exposure time obtained for NGC\,5170, these
issues may have played a role in suppressing any faint spurious signal
that could mimic diffuse X-ray emission around this galaxy. In
addition, we cannot exclude the possibility that the spurious halo
signal around NGC\,5746 had an additional contribution not present in
the NGC\,5170 data, e.g.\ from faint unresolved sources, such as the
one hinted at immediately east of the NGC\,5746 disk in
Figs.~\ref{fig,overlays} and \ref{fig,surfbright}a.

\subsection{Constraints on Halo Gas Properties}\label{sec,prop}

\begin{table*}
\begin{center}
\caption{Derived constraints on hot halo gas\label{tab,halos}}
\begin{tabular}{cccccccc}
  \tableline  \hline
  Galaxy & $T$ & $Z$ & $L_{\rm X}$ & $\langle n_e \rangle$ & $M_{\rm hot}$ & 
  $\langle t_{\rm cool}\rangle$ & $\langle \dot M \rangle $ \\
  & (keV) & (Z$_\odot$) & (erg~s$^{-1}$)& (cm$^{-3}$) & (M$_\odot$) & (Gyr) 
  & M$_\odot$~yr$^{-1}$ \\
  \tableline
  NGC\,5746 & 0.3 (fixed) & 0.2 (fixed) & $<4.0\times 10^{39}$ & 
  $<3.5\times 10^{-4}$ & $<2.3\times 10^9$ & $>7.9$ & $<0.3$ \\
  NGC\,5170 & 0.2 (fixed) & 0.2 (fixed) & $<7.5\times 10^{39}$ & 
  $<5.3\times 10^{-4}$ & $<3.4\times 10^9$ & $>2.3$ & $<1.5$ \\
  \tableline
\end{tabular}
\tablecomments{All constraints are given at $3\sigma$ significance and
  are derived within $r=40$~kpc, with the $D_{25}$ ellipses excluded.
  $L_{\rm X}$ is given in the 0.3--2~keV band, corrected for Galactic
  absorption, and $\langle \dot M \rangle \approx M_{\rm hot}/\langle
  t_{\rm cool} \rangle$ is the mean cool-out rate of hot gas within
  the assumed aperture.}
\end{center}
\end{table*}

Based on the absence of detectable extraplanar emission around either
of our targets, we can constrain some basic properties of any hot halo
gas surrounding these galaxies. In this context, it is important not
to confuse the signature of cooling halo gas with X-ray emission
related to processes in the disk, such as the integrated emission from
unresolved discrete sources and from hot gas heated by supernovae and
stellar winds. The optical disk of either galaxy is thus excluded in
the following analysis. On the other hand, the apparent absence of
significant X-ray, H$\alpha$, or radio emission outside the optical
disks, combined with the low specific star formation and supernova
rates of the galaxies as discussed in \S~\ref{sec,sample}, suggest
minimal contamination from outflows of hot gas driven beyond the
optical disk of either galaxy by stellar feedback.  Hence, we base our
constraints on the amount of accreted hot halo gas on the net count
rate limits evaluated inside the circular apertures underlying
Fig.~\ref{fig,surfbright}, i.e.\ a region extending to $r=40$~kpc from
immediately outside the $D_{25}$ ellipse of each galaxy.

In order to convert observed count rate limits into constraints on the
halo X-ray flux, we assume that the spectrum of any hot halo gas can
be described by a thermal plasma with a given temperature, metal
abundance, and volume filling factor (adopting the APEC model in {\sc
xspec} v11.3).  In the absence of direct observational information on
these quantities, we resort to more indirect observational constraints
and to simulation results.  For example, cosmological simulations
predict a remarkably tight correlation between the temperature of
infalling hot gas and $v_c$ of the disk (\S~\ref{sec,comparison}
below; see also \citealt{toft2002}), with $T$ being in good agreement
with a simple expectation for the virial temperature $T_{\rm
vir}\approx (1/2)\mu m_p v_c^2$ of the underlying dark matter halo. We
therefore nominally assume $T=T_{\rm vir} \approx 0.3$ and $\approx
0.2$~keV for any halo gas around NGC\,5746 and NGC\,5170,
respectively, based on the disk circular velocities listed in
Table~\ref{tab,galdata}. As further justification of this choice, we
note that gas much colder than $T_{\rm vir}$ would mainly emit at UV
wavelengths and likely be subject to rapid cool-out to $T\la 10^4$~K,
while gas much hotter would be subject to evaporation from the
galactic gravitational potential.  These situations are manifestly
distinct from the case of a gravitationally confined hot halo composed
of infalling shock-heated material as seen in our simulations, but it
is the amount and properties of any such material that is the focus of
our attention.

We further note that outside the galactic disks themselves, the
emission-weighted metal abundance of the X-ray gas in these
simulations displays a mean and standard deviation of
$Z=0.2\pm0.1$~Z$_\sun$, which is consistent with observational results
for the low-redshift intergalactic medium \citep{danf2006}, for the
outskirts of groups \citep{rasm2007} and clusters of galaxies (e.g.,
\citealt{tamu2004}), and for the intergalactic filaments from which
these structures accrete material \citep{fino2003}. Hence, for the
purpose of our spectral analysis, we here consider a fiducial value of
$Z=0.2$~Z$_\odot$ for the halo metal abundance (while noting that the
{\em mass-weighted} abundance in the simulations is substantially
lower, as discussed in \S~\ref{sec,temperatures}). Finally, for the
volume filling factor $\eta$ of hot gas, a value of unity is assumed
for simplicity, since our simulations (\S~\ref{sec,comparison})
indicate high values of $\eta \approx 0.8$--1.

Using these assumptions and the appropriate ACIS-I response files,
observed count rate limits were translated into corresponding
constraints on the APEC plasma model normalization $A$,
\begin{equation}\label{eq,xspec}
  A = \frac{10^{-14}}{4\pi D^2} \int{n_e n_H \mbox{ d}V},
\end{equation}
where $D$ is the assumed distance (Table~\ref{tab,galdata}), and $n_e$
and $n_{\rm H}$ are the number densities of electrons and hydrogen
atoms, respectively.  The integral represents the emission integral
$I_{\rm e} \equiv \int{n_e n_H \mbox{ d}V} \approx \eta n_e^2 V$ of
the hot halo gas inside the covered volume $V$. From the resulting
constraint on $I_{\rm e}$, we obtained upper limits to the mean hot
gas density $\langle n_e \rangle \sim (I_{\rm e}/V\eta)^{1/2} \approx
(I_{\rm e}/V)^{1/2}$ and total hot gas mass $M_{\rm hot} \sim m_p
\langle n_e \rangle V$ inside $V$. For $V$ itself, a spherical volume
of $r=40$ kpc was assumed, with the $D_{25}$ ``cylinder'' excluded
along the line of sight. Using the cooling curves $\Lambda(T,Z)$ of
\citet{suth1993}, we also derived constraints on the mean cooling time
$\langle t_{\rm cool} \rangle \approx 3kT/(\Lambda \langle n_e
\rangle)$ and cool-out rate $\langle \dot M \rangle \approx M_{\rm
hot}/\langle t_{\rm cool}\rangle$ of hot gas inside this volume.

Table~\ref{tab,halos} summarizes the derived constraints on halo
properties for both galaxies, where all values have been obtained from
the $3\sigma$ upper limits on the observed halo count rates using the
nominal values of $T$, $Z$, and $\eta$ discussed above.  Note that
although the two galaxies are roughly at the same distance, the
combination of shorter exposure time, higher Galactic absorbing
column, and assumed lower gas temperature of the NGC\,5170 halo, leads
to less tight constraints on the hot halo properties for this
galaxy. The lower temperature is particularly important, as {\em
Chandra} is not very sensitive to gas at $T\lesssim 0.2$~keV.

Although we have attempted to assume realistic values of $T$, $Z$, and
$\eta$ for the halo gas, it is worth briefly illustrating the impact
of these assumptions on the derived quantities. In particular, tighter
constraints would result for either galaxy if assuming a higher halo
temperature or metallicity. For our best constrained case of
NGC\,5746, increasing both the assumed $T$ and $Z$ by 50\% compared to
the adopted values in Table~\ref{tab,halos} would reduce our upper
limits on $L_{\rm X}$ and $M_{\rm hot}$ by roughly 30\%, while
doubling the lower limit on $\langle t_{\rm cool} \rangle$. For
NGC\,5170, the corresponding changes would be a 40\% reduction of
$L_{\rm X}$ and $M_{\rm hot}$, with $\langle t_{\rm cool} \rangle$
increasing by a factor 2.8. Also note that if employing uncertainties
at the 90\% confidence level rather than the more conservative
$3\sigma$ limits in Table~\ref{tab,halos}, we would instead obtain
$L_{\rm X}< 2.2\mbox{\,}(4.1)\times 10^{39}$~erg~s$^{-1}$ and $M_{\rm
hot} = 1.7\mbox{\,}(2.5) \times 10^9$~M$_\sun$ for NGC\,5746
(NGC\,5170). As will be apparent, none of these modifications would
affect our overall conclusions.

\section{Comparison to simulations of galaxy formation}\label{sec,comparison}

The lack of any detectable extraplanar hot gas related to stellar
feedback is not surprising for either of our target galaxies, given
the absence of any corresponding extraplanar H$\alpha$ or radio
emission. The derived constraint on the hot halo luminosity of
NGC\,5746, however, is also an order of magnitude below the naive
expectation from the \citet{toft2002} simulations for cosmologically
accreted gas around a galaxy of this circular velocity. Care should be
exercised in directly comparing the \citet{toft2002} predictions to
our results though, as the former also include the emission from
within the optical disk. Moreover, as mentioned in \S~\ref{sec,intro},
all the massive ($v_c \ga 250$~km~s$^{-1}$) galaxies in the
\citet{toft2002} study were formed in the old "standard" CDM cosmology
($\Omega_m=1$, $\Omega_{\Lambda}=0$), with an assumed cosmic baryon
fraction $f_b=0.05$ or 0.10, well below current estimates ($f_b=0.17$;
\citealt{sper2007}), and with no metal-dependence on the radiative
cooling~rate.  The aim of this section is to explore the predictions
of more sophisticated higher-resolution simulations assuming the
currently prevailing $\Lambda$CDM cosmology, and test whether these
can accommodate the observational constraints obtained for both our
galaxies.

\subsection{Cosmological Simulations}

For this purpose, we have employed cosmological TreeSPH simulations of
galaxy formation and evolution in a flat ($\Omega_m=0.3,
\Omega_\Lambda=0.7$) $\Lambda$CDM cosmology.  The code, described in
detail in \citet{somm2005} and \citet{rome2006}, incorporates
energetic stellar feedback in the form of starburst-driven galactic
winds. Contrary to our earlier simulations used by \citet{toft2002} to
predict halo properties (as discussed in \S~\ref{sec,intro} and
\S~\ref{sec,sample}), the present simulations also incorporate
self-consistent chemical evolution of the gas, including
non-instantaneous recycling. They have also been evolved in time
according to the entropy equation solving scheme of \citet{spri2002},
rather than being based on the thermal energy equation, a modification
which provides increased numerical accuracy in lower-resolution
regions. In addition, the present simulations assume a cosmic baryon
fraction of $f_b=0.15$ and have all been run from an initial redshift
of $z=39$ with a mass resolution eight times higher than those
described by \citet{toft2002}. At 0.5~Gyr before $z=0$, each hot gas
particle in the simulations was further split into 64 equal-mass
particles to provide additional resolution of the hot gas phase, a
procedure very similar to the one described by \citet{peek2008} based
on the same cosmological simulations.  The final gas-phase resolution
is thus 512 times that of the \citet{toft2002} simulations, with the
best resolved individual galaxies at $z=0$ featuring a total of $\sim
1.3\times 10^5$ hot gas particles within the central $r=100$~kpc, a
hot gas mass resolution of $\sim 1.7\times 10^4$~M$_{\sun}$, and an
SPH smoothing length of $h_{\rm SPH} \approx 1.5$~kpc. As discussed by
\citet{peek2008}, this resolution is sufficient to track thermal
instabilities in the hot halo gas that may lead to the formation of
warm ($T\sim 10^4$~K) clouds within the ambient halo.

Although the simulation code includes stellar feedback, optionally
enhanced to mimic AGN outflows, powerful starburst-driven winds only
occur in the simulations at early times ($z\ga 4$--5), where they help
to solve the angular momentum problem, the missing satellites problem,
and possibly other problems related to the cold dark matter scenario
(see, e.g., \citealt{somm2003}). At $z\la 4$, the SN~II feedback is
quite moderate in our simulations; for a typical $v_c
=220$~km~s$^{-1}$ disk galaxy, the star formation rate at $z=0$ is
$\sim 0.5-1$~M$_\odot$~yr$^{-1}$, comparable to the modest star
formation rate of the Milky Way disk.  The feedback of the individual
star particles in the simulations is furthermore spatially and
temporally uncorrelated (this is particularly true for SN~Ia, whose
mechanical, but not chemical, feedback is consequently neglected), so
galactic winds do not easily develop at low redshift (see also
\citealt*{dahl2006}). Hence, a comparison between the simulation
output at $z=0$ and our two relatively quiescent galaxies seems
justified.  It is worth emphasizing that the feedback strength in the
simulations is calibrated to reproduce the optical and morphological
properties of observed galaxies.  The X-ray emission is thus predicted
without any additional adjustable parameters.

For the calculation of X-ray properties of the simulated galaxies, we
followed the general procedure outlined by \citet{toft2002}. For each
of the simulated galaxies at $z= 0$, a catalog of SPH gas particle
positions, densities, temperatures, and masses was created.  Each SPH
particle was treated as an optically thin thermal plasma, and the
associated X-ray luminosity calculated at the relevant position in a
given photon energy band using the {\sc meka} plasma emissivity code
\citep*{mewe1986}. Total X-ray luminosities were then computed by
summation over all particles in the volume of interest. The greatly
increased numerical resolution of the present simulations at $z=0$
allowed us to bypass the additional smoothing over the individual gas
particles employed by \citet{toft2002}. However, as mentioned above,
the increased resolution also facilitates the formation of small
clouds within the hot halo via local thermal instabilities.  Since the
presence of such relatively dense and cold clouds may unphysically
boost the X-ray output of the hot gas in its immediate vicinity (due
to the local smoothing of particle properties inherent in SPH
techniques), hot gas particles within $h_{\rm SPH}$ of ``cold''
($T<3\times 10^4$~K) gas were discarded from the X-ray calculations.
The removed gas generally constitutes an insignificant fraction ($\la
1$\%) of the total hot gas mass considered.

A few comparisons between NGC\,5746 and results from an earlier
version of these simulations were performed in \cite{pede2006}. Here
we present a more detailed comparison based on the updated {\em
Chandra} analysis of NGC\,5746, while also including the results for
NGC\,5170 and for other, more actively star-forming, spirals as taken
from the literature. Further comparisons are made to our previous
results described in \citet{toft2002}, with the X-ray properties of
the latter here re-calculated for the adopted aperture and energy
band. However, in order to provide as fair a comparison as possible
between the two simulation samples, we only consider the subset of the
\citet{toft2002} galaxies that was formed formed in simulations
assuming the same flat ($\Omega_m=0.3$, $\Omega_{\Lambda}=0.7$)
$\Lambda$CDM cosmology as adopted in our new simulations, and which
assumed $f_b = 0.10$ (along with a fixed primordial chemical gas
composition). Note, as mentioned above, that this excludes all the
\citet{toft2002} galaxies with $v_c \geq 250$~km~s$^{-1}$, so a direct
comparison of the Toft et~al.\ results to our observed galaxies is not
attempted here.

\subsection{Hot Halo X-ray Luminosities}

One of the hot halo properties that can be most straightforwardly
compared to observations is the diffuse X-ray luminosity. For all
simulated galaxies, we derived the halo $L_{\rm X}$ within the
aperture also adopted for the NGC\,5746 halo analysis (i.e.\ a
spherical volume extending to $r=40$~kpc, with the $D_{25}$ ellipse
excluded as described in \S~\ref{sec,prop}). The results are presented
in Figure~\ref{fig,LxVc}, which shows a clear trend in hot halo
$L_{\rm X}$ with galaxy mass, measured by $v_c$, for the simulated
galaxies. Also shown are the observational constraints on the halo
luminosities of NGC\,5170 and NGC\,5746. The simulation results are
generally consistent with the constraints obtained for our observed
galaxies, and we conclude that, within the adopted apertures, the lack
of detectable extraplanar X-ray emission around either target agrees
with these updated predictions for emission from cosmologically
accreted gas.

\begin{figure}
\begin{center}
\epsscale{1.19}
\plotone{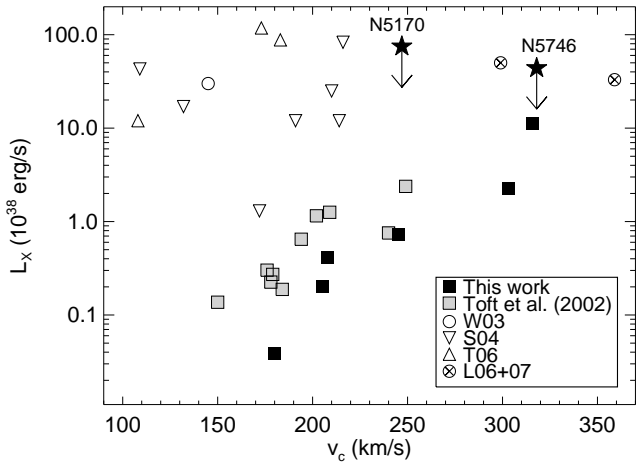}
\end{center}
\figcaption{0.3--2~keV X-ray luminosities versus $v_c$ for our
  simulated galaxies (black squares), and for those of Toft et~al.\
  (2002; gray squares), all calculated assuming the NGC\,5746 halo
  aperture. Also shown are observational results for extraplanar
  emission seen around other spirals, taken from
  \citet[``W03'']{wang2003}, \citet[``S04'']{stri2004a},
  \citet[``T06'']{tull2006a}, and \citet[``L06+07'']{li2006,li2007}.
\label{fig,LxVc}}
\end{figure}

There seems to be reasonable agreement between the results of
\citet{toft2002} and those of our newer simulations, but with a
tendency towards a somewhat lower $L_{\rm X}$ at a given $v_c$ for the
latter. For example, for Milky Way--sized galaxies ($v_c$ in the range
200--250~km~s$^{-1}$), the mean halo luminosity for the
\citet{toft2002} galaxies is $1.4\times 10^{38}$~erg~s$^{-1}$, 75\%
higher than the corresponding result for the more recent simulations.
We attribute this in part to the increased numerical resolution of the
latter, as discussed in more detail later. Nevertheless, the
pseudo-bolometric (0.01--10~keV) halo luminosities $L_{\rm X,bol}$ in
our new simulations remain consistent with the $L_{\rm X}\propto
v_c^{5-7}$ relationship found by \citet{toft2002}, with an orthogonal
regression fit in log\,$L_{\rm X}$--log\,$v_c$ space yielding $L_{\rm
X,bol} \propto v_c^{5.0\pm 0.7}$ for the new simulations. It is
interesting to note that this dependence on the characteristic stellar
velocity of the system mirrors that of the diffuse gas in {\em
early-type} galaxies, whose X-ray luminosity roughly obeys $L_{\rm
X}\propto L_B^2 \propto \sigma^{6-7}$ (based on the Faber-Jackson
relation and the $L_{\rm X}$--$L_B$ relation of \citealt{osul2001}).

The amount of hot X-ray gas surrounding the galaxies in our
simulations depends not only on the mass of each galaxy's dark matter
halo but also on the detailed accretion history of this
halo. Numerical calculations indicate that accretion histories of
galaxy-sized halos can be highly disparate even for isolated galaxies
of comparable mass (e.g.\ \citealt{vand2002,dero2008}). Hence, any
dependence of hot halo properties on $v_c$ can a priori be expected to
be subject to considerable scatter. Indeed, despite the clear trend in
$L_{\rm X}$ with $v_c$ for the simulated galaxies, it is important
also to emphasize the considerable variation in predicted halo $L_{\rm
X}$ at a given circular velocity in Fig.~\ref{fig,LxVc}.  For example,
the two most massive galaxies in our new simulation sample display a
factor of five variation in 0.3--2~keV $L_{\rm X}$ within the adopted
aperture and yet are separated in ``mass'' by only $\Delta v_c =
13$~km~s$^{-1}$, comparable to typical observational uncertainties on
$v_c$. This large variation at a given $v_c$ is clearly an issue to
keep in mind when comparing predicted and observed halo luminosities,
or when attempting to predict hot halo fluxes for actual galaxies on
the basis of simulation results.

In order to also facilitate comparison of the simulation predictions
to other observational results, we have compiled available constraints
from the literature on extraplanar emission seen around other spirals,
although we make no claims as to the completeness of this compilation
(note though that we have not included the results of
\citealt{sun2007}, because it is not entirely clear to what extent, if
any, these spirals show {\em extraplanar} emission). These results are
also included in Fig.~\ref{fig,LxVc}, but we stress that the halo
luminosities in these cases have not necessarily been derived within
the same aperture as adopted for NGC\,5746 and our simulated
galaxies. For example, the results of \citet{stri2004a} were all
derived at a distance $|z|>2$~kpc from the disk midplane. If instead
adopting this aperture, the increase in $L_{\rm X}$ predicted for our
simulated galaxies is at the 10--40~per~cent level, with no systematic
dependence on $v_c$. We expect similar corrections for the apertures
used for the additional galaxies in the \citet{tull2006a} sample,
where halo X-ray properties were constrained outside the H$\alpha$
disk, again typically at $|z|>2$~kpc from the midplane. These changes
are relatively small, well within the scatter seen for the simulated
galaxies.

The \citet{li2006,li2007} results in Fig.~\ref{fig,LxVc} include
unresolved emission within regions of the optical disk (though with
the prominent dust lane in the Sombrero galaxy excluded;
\citealt{li2007}). Hence, the true halo luminosities of these galaxies
should be somewhat lower than indicated in this figure, although the
apertures employed by Li et~al.\ extend to smaller radii than adopted
for NGC\,5746 and hence our simulated galaxies. We include these two
data points in Fig.~\ref{fig,LxVc} for completeness, in particular
because the result for the massive ($v_c \approx 359$~km~s$^{-1}$;
HyperLeda database) Sombrero galaxy clearly poses a strong constraint
on theoretical models, and since Fig.~\ref{fig,LxVc} suggests that the
impact of its $\sim 10^9$~M$_\sun$ supermassive black hole on the
diffuse $L_{\rm X}$ must be rather modest. We also note that a very
low halo X-ray luminosity of $3\times 10^{38}$~erg~s$^{-1}$ has been
reported for the $v_c \approx 250$~km~s$^{-1}$ Sb spiral NGC\,4565
\citep{wang2005}. This result would be broadly consistent with the new
simulation predictions in Fig.~\ref{fig,LxVc} if assuming similar
apertures. However, the result may be affected by emission from a
nearby background cluster of galaxies \citep{wang2005}, and the
details of the {\em Chandra} analysis have not been published, so we
have chosen not to include this data point in Fig.~\ref{fig,LxVc}.

Keeping the above caveats in mind, Fig.~\ref{fig,LxVc} indicates that
the theoretical predictions can generally be accommodated by the
observational results for the (presumably mainly starburst-generated)
hot halos seen in other studies. The latter show X-ray luminosities
for low- and intermediate-mass spirals which are up to three orders of
magnitude above the simulation results at a given $v_c$ but which are
also roughly independent on $v_c$ and show considerably larger scatter
at a given $v_c$. It is, nevertheless, interesting that X-ray
observations of high-$v_c$ spirals (e.g., \citealt{li2007}) are now
providing luminosity constraints which are comparable to simulation
predictions within the assumed aperture.

\subsection{Halo Gas Temperatures}\label{sec,temperatures}

The inferred temperature of the hot halo gas in observations and
simulations is a useful diagnostic of the nature and origin of this
gas. Figure~\ref{fig,TVc} shows the emission-weighted temperature
$\langle T \rangle$ of hot halo gas against $v_c$, for all gas within
the (100~kpc)$^3$ simulation volume surrounding each simulated
galaxy. The predictions for the halo temperature are generally very
similar to the expected virial temperature $T_{\rm vir} \propto
v_c^2$, as anticipated for accreting, shock-heated material, but with
a slight tendency for $\langle T\rangle$ to mildly exceed $T_{\rm
vir}$. A plausible explanation for this is that our estimate of
$\langle T\rangle$ is dominated by emission close to the disk, where
the gas temperature rises slightly above $T_{\rm vir}$, as
demonstrated below.  We note that a halo gas temperature slightly
exceeding the virial temperature of the dark matter is also a generic
feature of the phenomenological hot halo models considered by
\citet{fuku2006}.

\begin{figure}
 \epsscale{1.18} 
 \plotone{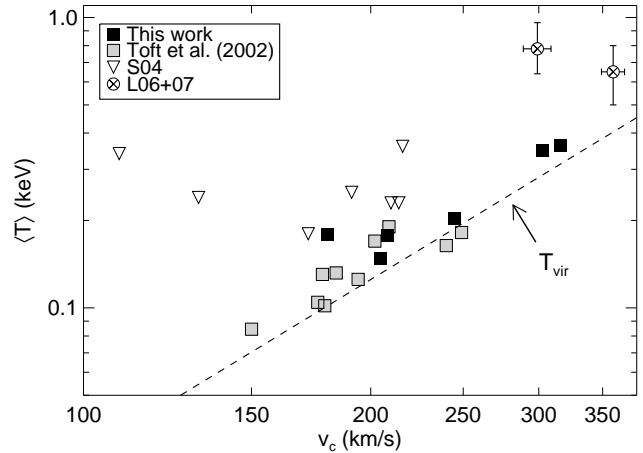}
\figcaption{Global emission-weighted hot halo temperature versus $v_c$
  for our simulated galaxies (squares), along with observational
  results for extraplanar emission around other spirals, with symbols
  as in Fig.~\ref{fig,LxVc}. Dashed line shows the virial estimate
  $T_{\rm vir}\approx (1/2)\mu m_p v_c^2$ of the hot halo temperature.
 \label{fig,TVc}}
\end{figure}

In contrast to the clear $T$--$v_c$ trend exhibited by the simulation
sample, published temperatures for the \citet{stri2004a} and
\citet{li2006,li2007} samples show no obvious systematic dependence on
$v_c$ (note that single emission-weighted halo temperatures are not
available for the samples of \citet{wang2003} or \citet{tull2006a}, so
these galaxies are not included in this plot). Any trend with $v_c$
for the observed galaxies must thus be subject to considerable
scatter, and is possibly modulated by a dependence on additional
parameters such as the star formation rate.  Also note the absence of
observed data points with $T \lesssim 0.2$~keV in the plot, which is
at least in part related to the difficulty of detecting such
low-temperature gas with {\em Chandra} or {\em XMM}. This also applies
to the very low--$v_c$ M82 --- excluded here for reasons of clarity
but included in Fig.~\ref{fig,strickland} --- for which
\citet{stri2004a} list separate temperatures of 0.27 and 0.37~keV for
the halo regions covered by ACIS-S and ACIS-I observations,
respectively.

For the galaxies with detected extraplanar emission, the lack of a
clear systematic trend in both $L_{\rm X}$ and $T$ with $v_c$, and
hence galactic mass, strongly argues against a gravitational origin
for this extraplanar emission. This interpretation gains support from
the facts that the hot halo temperatures of these galaxies
consistently exceed the virial temperatures expected from the circular
velocity of the disk and are furthermore invariant with respect to
$L_{\rm X}$ over four orders of magnitude in SFR \citep{grim2005}. In
general, the halo gas is almost certainly related to galactic wind or
fountain activity in these cases, rather than representing externally
accreted material. Indeed, unambiguous kinematic evidence for outflows
perpendicular to the disk has been uncovered in many of these galaxies
(\citealt{stri2004b} and references therein).

A single-temperature model was assumed when constraining hot halo
properties of our observed galaxies in \S~\ref{sec,prop}. To test the
validity of this assumption, we investigate spatial variations in the
predicted hot halo temperature in Fig.~\ref{fig,tprof}, in the form of
emission-weighted temperature profiles extracted perpendicular to the
disk. The reasonably uniform temperature distribution predicted
outside the disk suggests that our assumption of a single-$T$ model
for the observed galaxies is well motivated. The one exception is the
lowest-$v_c$ galaxy, which exhibits a very prominent warp in the {\em
cold} gas distribution that also affects the temperature distribution
of its hot gas, giving rise to the abrupt change seen around
$|z|\approx 25$~kpc. We note that a qualitatively very similar picture
to that in Fig.~\ref{fig,tprof} applies to the metal abundance of hot
halo gas, once again supporting the simplifying assumption of a single
hot gas phase adopted in \S~\ref{sec,prop}. This is true irrespective
of considering the emission-- or mass-weighted abundances, although
the latter is significantly lower within the adopted aperture, with a
mean and standard deviation of $Z=0.03\pm 0.01$~Z$_\sun$ compared to
the corresponding emission-weighted value of $0.2\pm 0.1$~Z$_\sun$.

\begin{figure}
\begin{center}
\epsscale{1.18}
\plotone{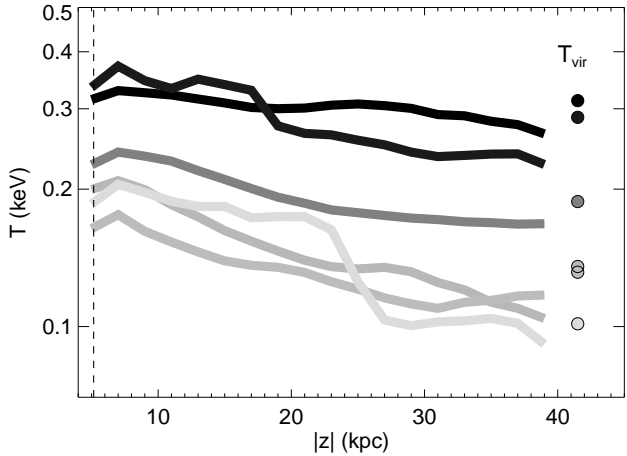}
\figcaption{Emission-weighted temperature profiles extracted
  perpendicular to the disk for the simulated galaxies. The gray-scale
  reflects the expected virial temperature $T_{\rm vir}$ (darker
  colors indicate higher $T_{\rm vir}$), with $T_{\rm vir}$ itself
  marked for each profile by a filled circle at an (arbitrary) large
  distance above the disk. Dashed vertical line marks the extent of
  the $D_{25}$ minor axis of NGC\,5746.
\label{fig,tprof}}
\end{center}
\end{figure}

Fig.~\ref{fig,tprof} also shows that the predicted halo temperature is
in good agreement with the virial temperature expected for the
simulated galaxies (i.e.\ $T\approx 0.3$~keV for a galaxy with
$v_c\approx 300$~km~s$^{-1}$) at intermediate distances ($|z|\sim
20$~kpc) from the disk, but drops slightly below this value further
out. The overall halo X-ray emission is dominated by gas closer to the
disk, where $T$ generally increases gently to slightly above $T_{\rm
vir}$, leading to the mildly super-virial global temperatures in
Fig.~\ref{fig,TVc}.  At even smaller distances from the disk midplane,
the temperature shows evidence of a beginning decline as a substantial
fraction of halo gas eventually reaches average densities which allow
for rapid cool-out.  This is reminiscent of a classical central
cooling flow, but it is worth emphasizing that cool-out of halo gas in
these simulations also proceeds via condensation of small
pressure-supported clouds at various distances from the disk, with a
monotonic increase in overall mass cool-out rate with decreasing
galactocentric distance \citep{somm2006, peek2008}. Also note that
although some gas is thus cooling rapidly in the central halo regions,
Fig.~\ref{fig,tprof} still suggests that the X-ray gas outside the
disk could appear near--isothermal. This behavior could potentially
compromise direct observational tests of whether halo gas is indeed
cooling to below X-ray temperatures (e.g.\ via X-ray grating
spectroscopy), as much of the cooling occurs very close to the optical
disk where absorption by cold gas and contamination from other X-ray
sources, diffuse and point-like, could be substantial.

\subsection{Hot Gas Mass and Cool-Out Rates}\label{sec,mass}

In the context of understanding the ``missing baryon'' problem in disk
galaxies (e.g.\ \citealt{mall2004,fuku2006,somm2006}), the total
amount of hot halo gas around such galaxies is a crucial quantity. For
our simulations, the correlation between hot gas mass $M_{\rm hot}$
within the NGC\,5746 aperture and $v_c$ is displayed in
Fig.~\ref{fig,Mgas}$a$. Given that the simulations also include cold,
non--X-ray emitting gas around the galaxies and that $M_{\rm hot}$ is
not an X-ray emission-weighted quantity, a low-temperature cut at
$T=0.05$~keV was imposed on the simulated galaxies in order to provide
a sensible comparison to the {\em Chandra} constraints on X-ray
emitting gas mass.  This temperature cut is well below $T_{\rm vir}$
for all the simulated galaxies (cf.\ Fig.~\ref{fig,TVc}), and is
probably close to the lower limit at which gas can be reliably
detected in emission in typical {\em Chandra} or {\em XMM} exposures
(see, for example, the discussion in \citealt{rasm2004b}).

\begin{figure}
\begin{center}
\epsscale{1.1}
\plotone{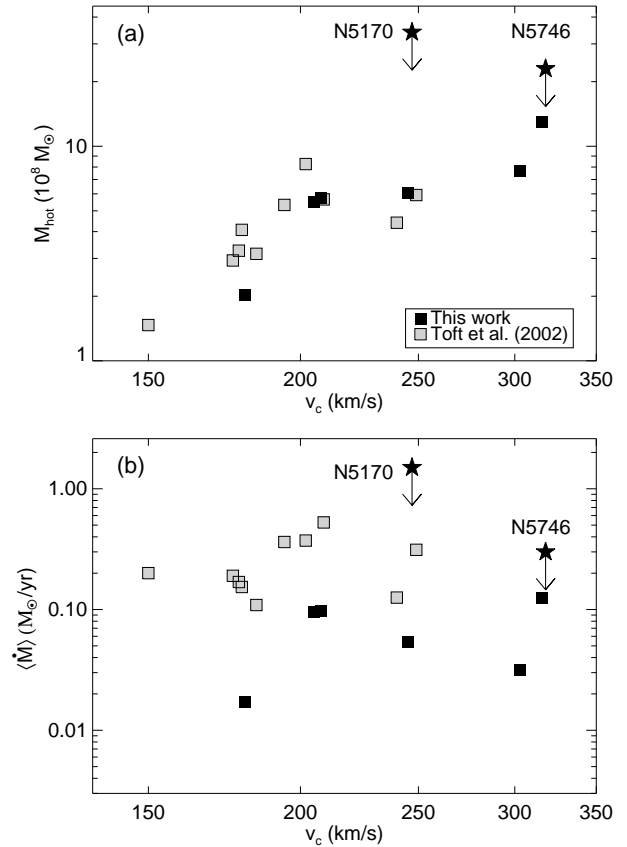}
\end{center}
\figcaption{(a) Mass of hot ($T\geq 0.05$~keV) gas as a function of
  $v_c$ within the NGC\,5746 halo aperture, with symbols as in
  Fig.~\ref{fig,LxVc}.  (b) The corresponding mean hot gas cool-out
  rate, $\langle \dot M \rangle \approx M_{\rm gas}/\langle t_{\rm
  cool}\rangle$.
\label{fig,Mgas}}
\end{figure}

As anticipated, a clear trend with $v_c$ is present for the simulation
results. These are also seen to be consistent with our observational
constraints, but we note that the $3\sigma$ upper limit derived for
NGC\,5746 is here only within a factor of 2--3 of the simulation
predictions for a galaxy of this mass. Clearly, any changes in the
assumed input physics in the simulations that would result in
significantly higher hot halo masses at $z=0$ are disfavored by our
observations. Also note that the overall hot gas masses within this
aperture at a given $v_c \ga 200$~km~s$^{-1}$ are similar for the two
simulation samples, despite the clear differences in corresponding
$L_{\rm X}$ visible in Fig.~\ref{fig,LxVc}. The main explanation is
that the increased numerical resolution employed here, and the removal
of a small amount of hot gas particles within an SPH smoothing length
of cold gas in the new simulations, reduces the amount of
(artificially) strongly X-ray emitting high-density halo gas.

The predicted amount of hot halo gas within the adopted aperture is
$M_{\rm hot} \la 10^9$~M$_\sun$ for Milky Way--sized galaxies,
corresponding to only $\la 2$\% of the baryonic disk mass of such
galaxies. The situation for our observed galaxies is unlikely to be
significantly different. In fact, \citet{rand2008} used the inferred
H{\sc i} rotation curve of NGC\,5746 to derive a dynamical mass inside
$r\approx 40$~kpc of $7.9\times 10^{11}$~M$_\sun$, which, combined
with $M_{\rm hot}$ from Table~\ref{tab,halos}, implies that any hot
halo gas outside the optical disk contributes less than 0.3\% of the
total mass budget within this radius. In terms of baryonic mass, the
inferred H{\sc i} mass in the disk of $1\times 10^{10}$~M$_\sun$,
added to a stellar mass of $1.6\times 10^{11}$~M$_\sun$ as estimated
from the optical parameters in Table~\ref{tab,galdata} using the
prescription of \citet{mann2005}, implies that $\la 1$\% of the
baryons in NGC\,5746 are in a hot phase within $r=40$~kpc. For
NGC\,5170, the corresponding constraint is $\la 4$\%, based on a
stellar mass of $6\times 10^{10}$~M$_\sun$ and an H{\sc i} mass of
$1.4\times 10^{10}$~M$_\sun$ \citep{bott1987}.

These fractions are expected to grow drastically with galactocentric
distance, however, as most of the diffuse hot baryons associated with
our simulated galaxies reside beyond the $r=40$~kpc aperture probed
here. As discussed in detail in \citet{somm2006}, the total mass of
hot halo gas within the virial radius may equal or exceed the mass of
cold baryons in the disk of a simulated $v_c\approx 300$~km~s$^{-1}$
galaxy. For our observed galaxies, this possibility is still allowed
by our constraints in Figure~\ref{fig,Mgas}, provided that the radial
distribution of any hot halo gas around our galaxies is not
significantly steeper than in these simulations. Deeper X-ray
observations would be required to test this scenario in detail, but
observational verification of this prediction would clearly help in
resolving the well-known ``missing baryon'' problem in disk galaxies.

As a final comparison between the simulations and our observational
constraints, Fig.~\ref{fig,Mgas}$b$ shows the results for the mean
cool-out rate $\langle \dot M \rangle \approx M_{\rm hot}/\langle
t_{\rm cool} \rangle$ of hot gas within the NGC\,5746 aperture.
Again, a temperature cut at $T=0.05$~keV was imposed on the
simulations, but $\dot M$ was otherwise estimated in the same manner
as for the observed galaxies (\S~\ref{sec,prop}), i.e.\ with hot gas
cooling times $\langle t_{\rm cool} \rangle$ derived assuming the
metal-dependent cooling curves $\Lambda(T,Z)$ of \cite{suth1993}. The
figure reveals substantial scatter at any given $v_c$ and no clear
systematic variation with this quantity. The simulation predictions
again remain consistent with our observational results as listed in
Table~\ref{tab,halos}. The difference between the \citet{toft2002} and
newer simulations is largely due to lower mean cooling times for the
hot gas in the former, for the reason discussed above.

It is important to note, however, that the estimated values of $\dot
M$ should not be identified with the actual accretion rate of hot halo
gas onto the disk, in part because these estimates do not include the
gas condensing out of the hot halo phase due to small-scale thermal
instabilities (see, e.g. \citealt{peek2008}), but more importantly
because our $\dot M$ does not take into account the gas cooling {\em
inside} the inner boundary of the adopted aperture. In practice, the
derived values of $\dot M$ should be considered lower limits to the
true disk accretion rates in the simulations.  For a detailed
discussion of actual accretion rates for the simulated galaxies, we
refer to \cite{somm2003} and \citet{peek2008}. Such a discussion is
beyond the scope of this work, also because it is not straightforward
to compare these rates to our observational constraints in
Table~\ref{tab,halos}.

\section{Discussion}\label{sec,discuss}

We emphasize that the observational part of this study was initiated
on the basis of the predictions of \citet{toft2002}, which suggested
that the performed observations of NGC\,5746 would be sufficiently
sensitive to reveal the presence of an accreted hot halo.  Our newer
simulations predict somewhat lower hot halo luminosities outside the
disk, in consistency with the inferred absence of detectable hot halo
gas around both NGC\,5746 and NGC\,5170. It is nevertheless clear that
observational results for massive galaxies are within factors of a few
of the theoretical predictions for halo luminosity and gas mass, and
can potentially begin to test the input physics in cosmological
simulations. For example, the predicted hot gas mass for $v_c \approx
300$~km~s$^{-1}$ galaxies is within a factor of 2--3 of the upper
limit derived for NGC\,5746. Numerical effects aside, inferred halo
X-ray luminosities of even more massive galaxies such as NGC\,4594
(cf.\ Fig.~\ref{fig,LxVc}) clearly leave limited scope for changes in
the simulation input physics that would raise the predicted
present-day halo $L_{\rm X}$, though with the caveat that the
simulations show some ``cosmic variance'' at a given $v_c$. Further
high-sensitivity observations may begin to seriously challenge the
theoretical models and provide useful constraints on, for example, the
galactic feedback prescriptions implemented in state-of-the-art
numerical work.

Compared to the earlier results presented by \citet{toft2002}, it is
also clear that the X-ray luminosity and cool-out rate of hot halo gas
are somewhat lower in the more recent simulations. One possible reason
is the higher mass resolution of the latter, which helps to suppress
the importance of a problem associated with the spatial smoothing
inherent in SPH techniques, namely the possibility that the properties
of hot gas particles can become strongly affected by those of much
colder and denser particles in their immediate vicinity. As a rough
test of the impact of improving the hot gas mass resolution, we
re-simulated our lowest-mass ($v_c\approx 180$~km~s$^{-1}$) galaxy
from $z=39$ at 64 times the \citet{toft2002} resolution, with each hot
gas particle then subsequently split into eight equal-mass ones at a
later stage. At high $z$, this version is thus eight times better
resolved than the one shown in Fig.~\ref{fig,LxVc}, and the halo
$L_{\rm X}$ is here found to be 50\% lower. An additional test for a
more massive galaxy, the $z\approx 1$ version of the $v_c\approx
245$~km~s$^{-1}$ disk in Fig.~\ref{fig,LxVc}, showed little change in
halo $L_{\rm X}$ at higher resolution, however. This suggests that
higher resolution may decrease the predicted $L_{\rm X}$ somewhat, but
the reduction is moderate (well within the factor 4--5 in ``cosmic
variance'' that is evident in Fig.~\ref{fig,LxVc}), and $L_{\rm X}$
may be close to convergence at the high-mass end.  We do not currently
have higher-resolution versions of our most massive galaxies at $z=0$
due to the prohibitive CPU times required, so more detailed tests of
the impact of numerical resolution will remain the subject of future
work.

Two other effects also act to reduce the present-day hot halo
luminosity in the updated simulations, namely the higher assumed
cosmic baryon fraction $f_b=0.15$ (vs.\ $f_b =0.10$ for the comparison
galaxies of \citealt{toft2002} considered here) and the higher metal
abundance resulting from implementing chemical evolution in the
simulation code. This may at first seem counter-intuitive, but the
increased radiative cooling efficiency implied by both these
modifications actually results in less hot and dense gas remaining in
the halo at $z=0$ for a given $v_c$ (see also \citealt{rasm2004a};
\citealt{somm2006}). Note that both the hot gas fraction and the total
``external'' (non--disk) baryon fraction inside the virial radius are
reduced at fixed $v_c$ when assuming higher $f_b$, not only because
the accreted shock-heated gas can leave the X-ray phase more rapidly,
but also because the virial radius itself decreases at a given $v_c$
as a consequence \citep{somm2006}.

Despite the reduced X-ray output in the current simulations, we
readily acknowledge the points raised by \citet{li2006,li2007} and
\citet{wang2007}. Their comparison of observational results obtained
for NGC\,2613 and NGC\,4594 to the 0.2--2 keV luminosities of
\citet{toft2002} indicate that the latter overpredict total (disk plus
halo) intrinsic X-ray luminosities by at least a factor of 10 for
these massive galaxies. This discrepancy is at least in part related
to the common over-cooling problem in hydrodynamical simulations, a
problem which also plagues the current simulations at some level
\citep{rome2006}. However, driven by the requirement to avoid
confusing any halo emission with emission produced by unrelated
processes, we have to some extent circumvented that issue here by only
considering the X-ray properties of gas outside the optical disk.
Note also that both the \citet{toft2002} and present simulations do
not incorporate the effects of intrinsic X-ray absorption by cold gas
and dust in the disk, which can naturally be substantial for edge-on
galaxies and may be non-trivial to properly correct for across the
full optical extent of the galaxy in observational studies.

While our results suggest that increased numerical resolution may help
to alleviate the issue of over-cooling somewhat, it is also possible,
as hypothesized by \citet{li2006,li2007}, that some over-cooling in
the simulations is related to incomplete treatment of feedback from
SN~Ia, if these can provide large-scale distributed heating. As
mentioned, {\em mechanical} feedback from SN~Ia is not included in the
present simulations, but \citet{wang2007} suggest that numerous SN~Ia
in early-type spirals (such as NGC\,2613, 4594, and 5746) could
generate a bulge wind which would be sufficiently powerful to suppress
the cooling of any infalling halo material and lower its X-ray
luminosity.  Observational verification of the presence of
SN~Ia--driven winds in spirals would certainly provide an important
new ingredient to be incorporated in numerical simulations. Our X-ray
data of NGC\,5746 and NGC\,5170 do not allow a direct observational
test of this scenario, apart from requiring such outflows to have
X-ray luminosities below the inferred halo upper limits. The
discussion pertaining to Fig.~\ref{fig,strickland}, although based on
somewhat simplistic assumptions, suggests that direct supernova
blow-out of disk material is unlikely to be important in our two
galaxies even if including SN~Ia. Our galaxies may therefore not
represent optimal targets for an observational test of this scenario.

\citet{wang2007} also proposes another possible solution to the
over-cooling problem, noting that the halo gas may be chemically
inhomogeneous but generally very low in metallicity. While metal-rich
clouds can cool out rapidly and fall towards the disk, most of the
halo gas may thus radiate inefficiently and would be difficult to
detect in either emission or absorption. Although an interesting
possibility, it is not obvious that this explanation may necessarily
help to solve the over-cooling problem. As discussed above, the
metal-free halos in the \citet{toft2002} simulations are actually {\em
more} X-ray luminous at a given $v_c$ at zero redshift than those in
our more recent simulations. Other factors such as the assumed cosmic
$f_b$ in the simulations may play a role for this result, as already
mentioned, but this difference between metal-enriched and metal-free
halos persists even when considering halos in simulations run with
identical cosmological parameters and numerical resolution
\citep{rasm2004a}. Allowing for chemical enrichment in our simulations
systematically reduces the predicted present-day halo luminosity for a
given $v_c$.

On the observational front, it is generally well established that disk
galaxies have experienced considerable stellar mass growth since $z
\approx 1$ (e.g.\ \citealt{bard2005} and references therein), and both
observational and theoretical evidence suggests that gas consumption,
rather than major merger activity, has played the dominant role in
driving this growth \citep{dera2008,dero2008,dadd2008}. The hot halos
discussed in this paper could potentially have supplied most of the
fuel required for the continued growth of disks, provided that
cool-out of such gas is not strongly inhibited by, for example, disk
feedback processes. Despite the absence of a detectable large-scale
X-ray corona around either of our target galaxies, hot coronal gas
surrounding both targets could still be present below our detection
limits. It remains unclear, however, to what extent any such gas is
currently cooling out to replenish the gas consumed by star formation
in their disks, and if so, whether this proceeds mainly as a central
cooling flow or through condensation and infall of small clouds across
a wide range of galactocentric distances within a hot halo. For
NGC\,5746, our naive constraint on the cool-out rate of hot halo gas
outside the optical disk, $\dot M < 0.3$~M$_\sun$~yr$^{-1}$, suggests
that large-scale cooling from a hot halo cannot balance the SFR in the
disk of $0.9\pm 0.2$~M$_\sun$~yr$^{-1}$. It is important to reiterate,
however, that the estimated cooling rates should be considered lower
limits to the total rate at which halo gas may leave the hot phase and
accrete onto the disk (\S~\ref{sec,mass}), so our results are not
necessarily inconsistent with the possibility that a hot halo around
NGC\,5746 can supply the disk with material at rates comparable to the
SFR.  For NGC\,5170, the weaker upper limit on $\dot M$ is a factor of
$\sim 3$ above the inferred SFR, again precluding any definitive
conclusions in this regard.

As suggested by a number of theoretical models, and, indeed, by the
simulations presented here (see \citealt{peek2008}), local thermal
instabilities within a hot halo can also produce small
pressure-supported clouds that would fall towards the disk on
time-scales well below the cooling time of the ambient hot gas. This
would contribute to feeding ongoing star formation and could
potentially represent the dominant accretion mode for Milky Way--sized
spirals \citep{mall2004}. However, H{\sc i} observations of nearby
spirals generally suggest that infall of cold gas does not occur at
rates sufficient to sustain star formation in the disk
\citep{sanc2008}. Hence, infall of additional, possibly hot, material
(at an average rate of $\approx 1$~M$_\sun$~yr$^{-1}$) seems required
to balance gas consumption by star formation, in line with the
findings of \citet{peek2008}. Of particular relevance here is the
study of \citet{rand2008}, who searched for evidence of an extended
H{\sc i} component around NGC\,5746 associated with infalling cold
clouds.  While the presence of vertically extended H{\sc i} was
detected, the majority of this gas is best interpreted as a warp
rather than as an extended H{\sc i} halo such as has been reported
around a number of other nearby spirals \citep{frat2006}. A number of
high-$|z|$ neutral clouds were also seen, possibly representing
infalling material, though association with a disk--halo flow cannot
be excluded for most of these. No corresponding study has, to our
knowledge, been conducted for NGC\,5170.

Thus, given the present data, it remains debatable whether the current
SFR in galaxies such as NGC\,5746 can be balanced by infall of halo
material. Cold gas accretion alone seems insufficient both in
NGC\,5746 and in other nearby spirals, and our upper limit on the hot
gas infall rate does not include gas cooling out within the disk
region itself, so the interpretation of the observational X-ray
results is not straightforward. It is apparent, however, that cooling
of hot halo gas outside the optical $D_{25}$ ellipse may not be able
to balance star formation in NGC\,5746. Interestingly,
\citet{frat2008} offer a different accretion scenario that may
reconcile the discrepant accretion and SFR estimates in nearby spirals
without invoking substantial direct cool-out of hot halo material. In
their model, accretion has an important contribution from
supernova--ejected fountain clouds which sweep up (hot or cold)
ambient halo gas before returning to the disk, losing angular momentum
in the process. Their combined fountain plus accretion models provide
a better match to the extraplanar H{\sc i} velocity distribution in a
few well-studied edge-on spirals than a fountain model alone, and
yield accretion rates comparable to the SFR in these galaxies. Direct
observational verification of this scenario may prove difficult,
however, and cosmological simulations still lack the resolution to
adequately capture such processes.

\section{Summary and Conclusions}\label{sec,summary}

We have used X-ray data to re-visit our earlier claimed detection
\citep{pede2006} of a large-scale diffuse X-ray halo surrounding the
massive edge-on spiral NGC\,5746. An analogous analysis of the less
massive NGC\,5170 has been completed for comparison. Our updated {\em
Chandra} analysis employs calibration data which accounts for the
positional dependence of the contaminant on the {\em Chandra} optical
blocking filters (unlike the \citealt{pede2006} analysis, which could
not take this into account), and has failed to reveal a statistically
significant detection of extraplanar X-ray emission around either
galaxy. We find that both the spatial and spectral hot halo properties
of NGC\,5746 as reported by \citet{pede2006} are consistent with being
due to the spatial variation of the degradation in the ACIS CCD
quantum efficiencies.

Both our target galaxies show no signs of significant nuclear or star
formation activity. This is supported by our H$\alpha$ imaging of both
galaxies and by radio data taken from the literature, suggesting that
the constraints on extraplanar X-ray emission obtained for both
galaxies can be used to also place constraints on the properties of
intergalactic, shock-heated material accreting onto their disks as
predicted by disk galaxy formation models and cosmological
hydro-simulations. If assuming halo gas close to the expected virial
temperature as suggested by such theoretical work, we constrain the
total 0.3--2~keV luminosity of the halos within $r=40$~kpc from the
optical disk to $L_{\rm X} < 4.0\times 10^{39}$ (NGC\,5746) and
$<7.5\times 10^{39}$~erg~s$^{-1}$ (NGC\,5170) at $3\sigma$
significance. The corresponding hot gas masses are $M_{\rm hot}
<2.3\times 10^9$ and $< 3.4\times 10^9$~M$_\sun$, respectively.

We have performed a detailed comparison of these constraints to
results of recent cosmological simulations of galaxy formation and
evolution. These simulations represent an update of our earlier work
in \citet{toft2002}, and predict hot halo luminosities and gas masses
which are consistent with the limits obtained for both NGC\,5746 and
NGC\,5170. Outside the optical disk, the simulations show hot halo
luminosities which are somewhat lower than the \citet{toft2002}
results, specifically by a factor of two for Milky Way--sized
galaxies, mainly due to increased numerical resolution, the assumption
of a higher baryon fraction, and the inclusion of chemical
feedback. We find that ``cosmic variance'' in simulated halo
properties can be quite substantial, with halo $L_{\rm X}$ potentially
varying by a factor 4--5 for galaxies of comparable mass as measured
by their disk circular velocity $v_c$.

Published results for all other galaxies with detected extraplanar
emission can generally accommodate our simulation predictions, but
observational results for high-mass spirals are clearly providing
strong constraints on theoretical models for disk galaxy
evolution. Based on the inferred halo temperatures and luminosities,
we confirm that the hot halo emission so far observed around other
spirals is in general predominantly related to activity in the disk
rather than being caused by infalling, externally accreted material.

An unambiguous detection of externally accreted X-ray emitting gas
around spirals is still pending. Deeper X-ray observations of
carefully selected targets are required to detect such halos, but,
encouragingly, our simulations suggest that this remains within reach
of current X-ray instrumentation. Based on a rather modest {\em
Chandra} exposure of less than 40~ks, we have constrained an X-ray
luminosity and hot gas mass around NGC\,5746 which are both within a
factor of a few of the simulation predictions for a galaxy of this
mass. Other existing observational constraints (e.g.,
\citealt{li2007}) are also close to the theoretical results and can
potentially begin to test the assumed input physics in numerical
simulations of galaxy formation and evolution.

\acknowledgments 

We thank the referee for useful comments that improved the clarity of
this paper.  This work has made use of the HyperLeda, NASA/IPAC (NED),
and Two Micron All Sky Survey (2MASS) extragalactic databases, and was
supported by the DFG Cluster of Excellence "Origin and Structure of
the Universe". JR acknowledges support provided by the National
Aeronautics and Space Administration through Chandra Postdoctoral
Fellowship Award Number PF7-80050 issued by the Chandra X-ray
Observatory Center, which is operated by the Smithsonian Astrophysical
Observatory for and on behalf of the National Aeronautics and Space
Administration under contract NAS8-03060. KP acknowledges support from
Instrument Center for Danish Astrophysics.  We gratefully acknowledge
abundant access to the computing facilities provided by the Danish
Centre for Scientific Computing (DCSC), with which all computations
reported in this paper were performed. The Dark Cosmology Centre is
funded by the Danish National Research Foundation.


\end{document}